\documentclass[12pt]{article}
\usepackage[utf8]{inputenc}
\usepackage{graphicx} 
\usepackage{dcolumn} 
\usepackage{bm} 
\usepackage{amsfonts}
\usepackage{amsthm}
\usepackage{amsmath}
\usepackage{amssymb}
\usepackage{epsfig}
\usepackage{color}
\usepackage{textcomp}
\usepackage{hyperref}
\usepackage{csquotes}
\usepackage{titlesec}
\usepackage{slashed}
\usepackage{caption}
\usepackage{subcaption}
\usepackage{booktabs}
\usepackage{multirow}
\usepackage{cite}
\usepackage{placeins}
\usepackage{comment}
\usepackage{braket}
\usepackage{authblk}
\usepackage{physics}
\usepackage{mathtools}
\numberwithin{equation}{section}

\title{\textbf{Quantum Inequalities from the Second Law}}
\author[1]{Andrea Palessandro \thanks{andrea.palessandro@gmail.com}}
\affil[1]{\small Società Italiana di Fisica}
\date{}

\begin{document}

\maketitle
\begin{abstract}
\noindent Quantum field theory allows local violations of the classical energy conditions, such as regions of negative energy density. The Ford--Roman quantum inequalities quantify how negative these regions can be and for how long. We present a derivation of these inequalities from the second law of thermodynamics, in an operational framework that couples the field to a small thermal detector. The detector acts as a physical probe of the field's energy density and, by the non-negativity of the entanglement entropy between the two systems, one can show that its heat loss is bounded from below by a state-independent quantity.
\end{abstract}

\section{Introduction}

Classical field theory enforces strict positivity properties on the stress tensor. For example, the local energy density $T_{00}$ of a classical scalar field is always nonnegative. Quantum field theory violates these pointwise energy conditions. Well known examples such as the Casimir effect \cite{Casimir}, squeezed states \cite{Walls, Wu} or moving mirrors \cite{Fulling} reveal that the renormalized energy density can be negative in finite regions of spacetime. Such violations are not arbitrary, however. A series of results beginning with Ford and Roman established that negative energy comes with quantitative trade-offs \cite{Ford1978,Ford1991}. Their quantum inequalities show that the time-averaged energy density measured along any worldline is bounded from below by a negative quantity that scales like the inverse fourth power of the sampling time. These bounds are universal in the sense that they hold for all states of the field, and they place important restrictions on exotic phenomena that rely on negative energy, such as warp drives, wormholes or violations of the second law \cite{Ford1, Ford2, Ford3, Ford4, Ford5}.

Most derivations of quantum inequalities proceed from field theoretic considerations. They rely on properties of the Wightman function, microlocal spectrum conditions or the positivity of certain quadratic forms \cite{FordRoman1995,FordRoman1997,Flanagan1997,PfenningFord1997PRD,PfenningFord1998PRD,FordPfenningRoman1998,FewsterEveson1998,Fewster2000,FewsterTeo2000,FewsterVerch2003,Fewster2007}. Ford originally motivated the existence of such bounds through thermodynamics, noting that arbitrarily large negative energy fluxes would allow violations of the second law. A rigorous connection between negative energy and thermodynamic constraints has been less explored.

In this work we develop an operational version of Ford's idea. Rather than treating the energy density as an observable to be inserted directly into the inequality, we couple the quantum field to a small thermal system that plays the role of a physical detector or thermometer. The detector interacts with the field for a finite time through a smooth switching function. During this interaction it may absorb or lose heat depending on the field state. The detector therefore provides an operational probe of the field's energy density through its heat exchange.

The key point is that the detector and the field evolve unitarily as a closed system, but we track only their reduced states. The entanglement generated between them is nonnegative, and this property implies a second law type constraint on the coarse grained entropies. To leading nontrivial order in the interaction strength, the entropy balance yields a state independent lower bound on the heat lost by the detector. Because this heat is a linear functional of the symmetric part of the field's two point function smeared along the detector's worldline, the resulting bound can be interpreted as a quantum inequality for a nonlocal, operationally defined probe of the energy density.

Our result therefore provides a thermodynamic route to bounds reminiscent of the Ford--Roman inequalities. It shows that negative energy cannot be made arbitrarily large or last arbitrarily long when measured by a physical thermometer. Instead, the detector’s heat loss is constrained by a universal lower bound that depends only on its own characteristics, such as its frequency, temperature and switching profile. This establishes a direct connection between the second law and the limitations on negative energy in quantum field theory within a fully operational framework.

\section{The setup}

Consider an inertial observer at $\mathbf{x}=0$ measuring the energy density of a free, massless scalar field $\phi$ in four-dimensional Minkowski spacetime. The energy density operator is
\begin{equation}
    T_{00}(t,\mathbf{0}) = \frac{1}{2}\!\left[(\partial_t \phi)^2 + (\nabla \phi)^2\right].
    \label{T00}
\end{equation}
Instead of measuring $T_{00}$ instantaneously, the observer averages it over time with a smooth, real sampling function $g(t)$ of width $\tau$:
\begin{equation}
    \mathcal{E}_g[\psi] = \int_{-\infty}^{\infty} dt\, g(t)^2\, \langle T_{00}(t,\mathbf{0}) \rangle_\psi,
    \label{Eg}
\end{equation}
where $\psi$ is an arbitrary state of the field $\phi$.
The function $g(t)$, typically normalized such that $\int g(t)^2 dt = 1$, represents the finite duration of the measurement. 
In the usual formulation of quantum inequalities one seeks a state independent lower bound on $\mathcal{E}_g[\psi]$ valid for all states $\psi$.

In this work we take a slightly different point of view. Rather than treating $\mathcal{E}_g[\psi]$ as a primitive observable, we introduce a small thermal system A that plays the role of a physical detector. The detector follows the inertial worldline $\mathbf{x}=0$ and couples locally to the field via a switching function $g(t)$. During the interaction the detector can either absorb or lose energy, and its mean heat exchange $\langle Q \rangle_\psi$ defines a measurable quantity that depends on the field state. We will look for a bound on $\langle Q \rangle_\psi$ obtained from entropy balance.

The detector is modeled as a harmonic oscillator of frequency $\Omega$,
\begin{equation}
    H_A = \frac{1}{2}(P_A^2 + \Omega^2 X_A^2),
    \label{H_A}
\end{equation}
initially in a thermal state at temperature $T = 1/\beta$:
\begin{equation}
    \rho_A^{(\beta)} = \frac{e^{-\beta H_A}}{Z_A}.
    \label{rhoA}
\end{equation}
The interaction Hamiltonian is taken to be of Unruh--DeWitt form \cite{Unruh, DeWitt},
\begin{equation}
    H_I(t) = \lambda\, g(t)\, \phi(t,\mathbf{0})\, X_A \equiv \lambda V(t),
\end{equation}
with small coupling $\lambda$. Physically, the thermometer couples to the field for a short time. The interaction strength rises and falls smoothly as $g(t)$. Afterward, the thermometer’s energy might have gone up or down depending on whether it absorbed or emitted energy into the field.

The total system evolves under $H_{\phi} + H_A + H_I(t)$ from an initial product state $\rho_0 = \rho_\psi \otimes \rho_A$\footnote{It will be understood that $\rho_A$ is the initial state (\ref{rhoA}).}. After the interaction, the field and the detector get slightly entangled. If the field has positive energy, the detector tends to absorb energy (heats up). If the field has negative energy, the detector tends to lose energy (cools down). So by measuring how the detector’s energy changes, we are effectively measuring the sampled energy density of the field.

The interaction picture evolution operator is 
\begin{equation}
    U_I(t_f) = \exp\left( -i \lambda \int_{-\infty}^{t_f} dt \ V(t) \right).
    \label{U_I}
\end{equation}
To second order in $\lambda$, (\ref{U_I}) is
\begin{equation}
    U_I = 1 - i \lambda \int dt \ V(t) - \lambda^2 \int dt \int dt' \ \theta(t-t') V(t) V(t') + O(\lambda^3). 
\end{equation}
The evolved final state is $\rho_f = U_I(t_f) \rho_0 U_I^\dagger(t_f)$. Hence,
\begin{equation}
    \rho_f = \rho_0 - i \lambda \int dt [V(t), \rho_0] - \lambda^2 \int dt \int dt' \theta(t-t') [V(t), [V(t'),\rho_0]] + O(\lambda^3).
    \label{rho_f}
\end{equation}

The total entropy can be defined as
\begin{equation}
    S_{\text{tot}}(t) = - \Tr_{\phi,A} \left[ \rho(t) \log \rho(t)\right].
\end{equation}
Note a general identity: for any positive operator $\rho$ and any unitary $U$\footnote{%
For any matrix $A$ and invertible $X$, and any function $f$ analytic on a domain containing the spectrum of $A$, one has
$f(XAX^{-1}) = X f(A) X^{-1}$.
Indeed, analyticity implies that $f$ admits a convergent power series
$f(z)=\sum_n c_n z^n$ on that domain, and hence
$f(A)=\sum_n c_n A^n$.
Since $(XAX^{-1})^n = X A^n X^{-1}$ for all $n\ge 0$, it follows that
$f(XAX^{-1})=\sum_n c_n (XAX^{-1})^n
= X \bigl(\sum_n c_n A^n\bigr) X^{-1}
= X f(A) X^{-1}$.}
,
\begin{equation}
    \log\left(U \rho U^\dagger\right) = U (\log \rho) U^\dagger,
\end{equation}
so the final entropy is\footnote{If not specified, the trace is understood to be taken over all detector + system degrees of freedom.}
\begin{equation}
\begin{aligned}
    S_{\text{tot}}(t_f)
        &= - \Tr[\rho_f \log \rho_f] \\
        &= - \Tr[U \rho_0 U^\dagger \log(U\rho_0 U^\dagger)] \\
        &= - \Tr[U \rho_0 U^\dagger U \log \rho_0 U^\dagger] \\
        &= - \Tr[\rho_0 \log \rho_0] \\
        &= S_{\text{tot}}(0).
        \label{entropy_conserved}
\end{aligned}
\end{equation}
The closed system $\phi+A$ evolves unitarily, so the total von Neumann entropy is conserved.

Even though the total entropy is fixed, the detector and the field individually evolve nonunitarily, so their entropies change. 

\section{The entropy of the detector}
Take for example the detector's reduced state $\rho_A(t) = \mathrm{Tr}_{\phi} \rho(t)$. The entropy of the detector is
\begin{equation}
    S_A(t) = - \Tr_A\left[ \rho_A(t) \log \rho_A(t)\right].
\end{equation}
The entropy change of the detector then is
\begin{equation}
    \Delta S_A = S(\rho_A(t_f)) - S(\rho_A).
\end{equation}
Let's define $\delta \rho_A \equiv \rho_A(t_f) - \rho_A$. Then we can write the entropy change as
\begin{equation}
    \Delta S_A = - \Tr_A\left[ (\rho_A+\delta \rho_A) \log(\rho_A + \delta \rho_A)\right] + \Tr_A\left[ \rho_A \log \rho_A\right].
    \label{deltaS}
\end{equation}
We expand $\log(\rho + \delta \rho)$\footnote{Remember that $\rho$ and $\delta \rho$ are operators.}:
\begin{equation}
    \log(\rho + \delta \rho) = \log \rho + \int_0^1 dt \rho^{-t} \delta \rho \rho^{t-1} + O(\delta \rho^2).
\end{equation}
Defining $L(\delta \rho) \equiv \int_0^1 dt \rho^{-t} \delta \rho \rho^{t-1}$ and substituting into (\ref{deltaS}) we get
\begin{equation}
    \Delta S_A = - \Tr_A\left[\rho_A L(\delta \rho_A)\right] -  \Tr_A\left[\delta \rho_A \log \rho_A\right] -  \Tr_A\left[\delta \rho_A L(\delta \rho_A)\right] + O(\delta \rho_A^3). 
\end{equation}
Let's focus on the first term. This is 
\begin{equation}
    \int_0^1 dt \Tr_A\left[ \rho_A \rho_A^{-t} \delta \rho_A \rho_A^{t-1}\right] = \int_0^1 dt \Tr_A\left[ \rho_A^{1-t} \delta \rho_A \rho_A^{t-1}\right] = \int_0^1 dt \Tr_A\left[ \delta \rho_A \right] = 0,
\end{equation}
where the second equality follows from the cyclicity of the trace, and the third from the fact that we only consider variations that preserve the trace, so that $\Tr(\rho) = 1$ and $\Tr(\delta \rho) = 0$ for all states $\rho$. Finally then
\begin{equation}
    \Delta S_A = - \Tr_A\left[\delta \rho_A \log \rho_A\right] - \Tr_A \left[ \delta \rho_A \int_0^1 dt  \rho_A^{-t} \delta \rho_A \rho_A^{t-1} \right] + O(\delta \rho_A^3).
    \label{deltaS_A}
\end{equation}
Now, from (\ref{rhoA})
\begin{equation}
    \log \rho_A= - \beta H_A-\log Z,
\end{equation}
and since $\Tr_A(\delta \rho_A) = 0$, 
\begin{equation}
    - \Tr[\delta \rho_A \log \rho_A]  = \beta \Tr[H_A \delta \rho_A] \equiv \beta \langle Q \rangle,
\end{equation}
where $\langle Q \rangle$ is the average energy gained (or lost) by the detector. This is
\begin{equation}
    \langle Q \rangle = \Tr\left[H_A(\rho_A(t_f)-\rho_A)\right].
    \label{Q}
\end{equation}
The second term in (\ref{deltaS_A}) is
\begin{equation}
   \delta S^{(2)} = - \Tr_A \left[ \delta \rho_A \int_0^1 dt  \rho_A^{-t} \delta \rho_A \rho_A^{t-1} \right].
   \label{deltaS2}
\end{equation}
For a thermal harmonic oscillator, we can diagonalize in the energy basis
\begin{equation}
    \rho_A = \sum_n \rho_n \ket{n} \bra{n}, \quad \rho_n = (1-e^{-\beta \Omega}) e^{-n\beta \Omega},
\end{equation}
and
\begin{equation}
    \delta \rho_A = \sum_{n,m} \delta \rho_{nm} \ket{n} \bra{m}.
\end{equation}
Insert into the trace (\ref{deltaS2}):
\begin{equation}
    \delta S^{(2)} = - \sum_{n,m} \left(\int_0^1 dt \rho_n^{-t} \rho_m^{t-1}\right) |\delta \rho_{nm}|^2.
\end{equation}
We compute the integral
\begin{equation}
    \int_0^1 dt \rho_n^{-t} \rho_m^{t-1} = \frac{\rho_n - \rho_m}{\rho_n \rho_m \left( \log \rho_n - \log \rho_m\right)}.
\end{equation}
Thus
\begin{equation}
    \delta S^{(2)} = - \sum_{n, m} \frac{\rho_n - \rho_m}{\rho_n \rho_m \left( \log \rho_n - \log \rho_m\right)} |\delta \rho_{nm}|^2,
    \label{deltaS2'}
\end{equation}
with the understanding that when $n=m$ the fraction is taken as $\rho_n^{-1}$.
Given that $\rho_n \geq 0 \ \forall n$, the fraction in (\ref{deltaS2'}) is always positive, meaning that $\delta S^{(2)} \leq 0$. In other words, the entropy is concave. This is just a restatement of Clausius inequality\footnote{The detector is initially in a thermal state which maximizes the von Neumann entropy for the given Hamiltonian. A thermal state is a local maximum  of the entropy functional, so small perturbations necessarily produce a second-order entropy change that is non-positive.}
, i.e. 
\begin{equation}
    \Delta S_A \leq \beta \langle Q \rangle.
\end{equation}
Let's calculate (\ref{deltaS2'}). The density perturbation is $\delta\rho_A = \Tr_\phi\left[ \rho_f - \rho_0\right]$. From (\ref{rho_f}) this is 
\begin{equation}
    \delta\rho_A = - i \lambda \int dt \Tr_\phi [V(t), \rho_0] + O(\lambda^2).
\end{equation}
The first term is 
\begin{equation}
    - i \lambda \int dt \Tr_\phi [V(t), \rho_0] = - i \lambda \int dt g(t) \langle \phi(t)\rangle_\psi [X_A(t), \rho_A].
\end{equation}
However, for any physically relevant state we have $\langle \phi(t)\rangle_\psi = 0$, therefore $ \delta\rho^{(1)}_A  = 0$ to first order. This means that the leading nonzero contribution to $\delta\rho_A$ is the second order term, $\delta\rho^{(2)}_A \sim \lambda^2$. 

Thus from (\ref{deltaS_A}), $\Delta S_A = \beta \langle Q \rangle + O(\lambda^4)$. To leading order then $\Delta S_A = \beta \langle Q \rangle$.

\section{The heat absorbed by the detector}
\noindent We want to calculate (\ref{Q}). Note that 
\begin{equation}
   \langle Q \rangle = \Tr_A\left[H_A(\rho_A(t_f)-\rho_A)\right] = \Tr_A\left[H_A(\Tr_\phi \rho_f-\Tr_\phi \rho_0)\right] = \Tr \left[ H_A(\rho_f - \rho_0)\right]. 
\end{equation}
This is because the operator $H_A \equiv H_A \otimes \mathbf{1}$ acts trivially on the field's degrees of freedom.

From (\ref{rho_f}), the first order term is
\begin{equation}
    - i \lambda \int dt \Tr \left[ H_A [V(t),\rho_0]\right] = - i \lambda \int dt \Tr \left[ [H_A,V(t)] \rho_0 \right],
\end{equation}
where the equality follows from the cyclicity of the trace. Since $[H_A,\phi(t)] = 0$, 
\begin{equation}
    [H_A,V(t)] = g(t) \phi(t) [H_A,X_A].
\end{equation}
But from (\ref{H_A}) it follows that $[H_A,X_A]= -i \dot{X}_A$. In a thermal state $\Tr [\rho_A^{(\beta)}\dot{X}_A] = 0$ so this linear term vanishes. Therefore, the leading contribution is second order:
\begin{equation}
    \langle Q \rangle = -\lambda^2 \int dt \int dt' \theta(t-t') \Tr \left[ H_A [V(t), [V(t'),\rho_0]]\right] + O(\lambda^3).
    \label{deltaE}
\end{equation}
We can use the cyclicity of the trace again to move $H_A$ inside the nested commutator:
\begin{equation}
    \Tr \left[ H_A [V, [V', \rho_0]]\right] = - \Tr \left[ \rho_0 [V', [H_A, V]]\right].
\end{equation}
Thus (\ref{deltaE}) becomes
\begin{equation}
    \langle Q \rangle = \lambda^2 \int dt \int dt' \theta(t-t') \Tr \left[ \rho_0 [V(t'), [H_A, V(t)]]\right].
    \label{deltaE'}
\end{equation}
We compute the inner commutator as above
\begin{equation}
    [H_A, V(t)] = g(t) \phi(t) [H_A, X_A] = -i g(t) \phi(t) \dot{X}_A(t).
\end{equation}
Therefore
\begin{equation}
    [V(t'), [H_A, V(t)]] = -i g(t) g(t') \left( \phi(t') X_A(t') \phi(t) \dot{X}_A(t)-\phi(t)\dot{X}_A(t) \phi(t') X_A(t')\right).
\end{equation}
Taking the trace over the product state $\rho_0 = \rho_\psi \otimes \rho_A$ in (\ref{deltaE'}) factorizes field and detector parts:
\begin{equation}
    \Tr \left[ \rho_0 \phi(t') X_A(t') \phi(t) \dot{X}_A(t)\right] = \Tr_\phi \left[ \rho_\psi \phi(t') \phi(t) \right] \Tr_A \left[ \rho_A  X_A(t') \dot{X}_A(t) \right].
\end{equation}

We now define the Wightman function on the worldline as
\begin{equation}
    W_\psi(t-t') =  \langle \phi(t) \phi(t') \rangle_\psi \equiv \Tr_\phi \left[ \rho_\psi \phi(t) \phi(t')\right],
\end{equation}
and the detector susceptibility as
\begin{equation}
    \chi_A(t-t') \equiv i \theta(t-t') \Tr_A\left[ \rho_A[X_A(t), X_A(t')]\right].
    \label{chi}
\end{equation}
With these definitions, (\ref{deltaE'}) can be written as
\begin{equation}
    \langle Q \rangle = \lambda^2 \int dt\,dt'\, g(t)\,g(t')\, \dot{\chi}_A(t-t')\, W_\psi(t-t').
    \label{Qangle}
\end{equation}
This is the mean energy absorbed by the detector.

\section{The entropy of the scalar field}
\noindent Similarly for the field $\phi$, take $\rho_\phi(t) = \Tr_A \rho(t)$. Then the field entropy is 
\begin{equation}
    S_\phi (t) = - \Tr_\phi [\rho_\phi(t) \log \rho_\phi(t)].
\end{equation}
Assuming the initial state $\rho_\phi \equiv \ket{\psi} \bra{\psi}$ is a pure state, 
\begin{equation}
    \rho_\phi(t_f) = \ket{\psi} \bra{\psi} + \delta \rho_\phi.
    \label{rhophi}
\end{equation}
Given that the entropy of a pure state is zero, the change in entropy is simply
\begin{equation}
    \Delta S_\phi = S_\phi (t_f) = - \Tr_\phi[\rho_\phi (t_f) \log \rho_\phi(t_f)].
    \label{deltaS_phi}
\end{equation}
Because the full evolution is trace preserving, we can write (\ref{rhophi}) as $\rho_\phi(t_f) = \ket{\psi} \bra{\psi} + \lambda^2 \sigma$, with $\Tr(\sigma)= 0$\footnote{Note that the formula (\ref{deltaS_A}) is not valid around a pure state. The Taylor expansion can only be used for a full-rank (strictly positive) density matrix.}.

Now we can decompose $\sigma$ with respect to the projector onto $\ket{\psi}$:
\begin{equation}
    P \equiv \ket{\psi} \bra{\psi}, \quad Q \equiv 1 - P.
\end{equation}
Then, 
\begin{equation}
    \sigma = P \sigma P + P\sigma Q + Q \sigma P + Q \sigma Q.
\end{equation}
Now let
\begin{equation}
    \alpha \equiv \bra{\psi}\sigma\ket{\psi} = \Tr(P \sigma P), \quad C\equiv Q \sigma Q, \quad B \equiv Q \sigma P.
    \label{alphadef}
\end{equation}
Then one can write $\sigma$ as
\begin{equation}
    \sigma = 
    \begin{pmatrix}
    \alpha & B^\dagger \\
    B & C
    \end{pmatrix}.
\end{equation}
The trace condition $\Tr(\sigma) = 0$ gives
\begin{equation}
    \alpha + \Tr(C) = 0.
\end{equation}
Therefore we can write (\ref{rhophi}) as
\begin{equation}
     \rho_\phi(t_f) = 
     \begin{pmatrix}
    1 + \lambda^2 \alpha & \lambda^2 B^\dagger \\
    \lambda^2 B & \lambda^2C
    \end{pmatrix} + O(\lambda^3).
    \label{rho_phi}
\end{equation}
We now find the eigenvalues of (\ref{rho_phi}) perturbatively in $\lambda$. To leading order, they are 
\begin{equation}
    p_0 = 1 + \lambda^2 \alpha + O(\lambda^4), \quad p_i = \lambda^2 \mu_i + O(\lambda^4),
\end{equation}
where $\mu_i$ are the eigenvalues of $C$, with $\sum_i \mu_i = -\alpha$\footnote{This ensures that $p_0 + \sum_i p_i = 1$ as imposed by unitarity.}.

Therefore, we can write the entropy (\ref{deltaS_phi}) as
\begin{equation}
    \Delta S_\phi = - \sum_k p_k \log p_k = - p_0 \log p_0 - \sum_{i \geq 1} p_i \log p_i.
\end{equation}
Again, to leading order in $\lambda$, this is 
\begin{equation}
    \Delta S_\phi = \lambda^2 \alpha (\log \lambda^2 - 1) - \lambda^2 \sum_{i \geq 1} \mu_i \log \mu_i + O(\lambda^4).
\end{equation}
The dominant behavior for small $\lambda$ is the non-analytic term\footnote{Entropy is not analytic around pure states, as these have zero eigenvalues.}
\begin{equation}
    \Delta S_\phi \sim \lambda^2 \alpha \log \lambda^2,
\end{equation}
where $\alpha = - \Tr(Q \sigma Q) \leq 0$.

From (\ref{rho_f}), 
\begin{equation}
    \delta \rho = - \lambda^2 \int dt dt' \theta(t-t') [V(t), [V(t'), \rho_0]] + O(\lambda^3).
\end{equation}
The reduced state of the field is $\rho_\phi = \Tr_A \rho$. Therefore, to leading order in $\lambda$
\begin{equation}
    \delta \rho_\phi = -\lambda^2 \int dt dt' \theta(t-t') \Tr_A[[V(t), [V(t'), \rho_0]]] + O(\lambda^3).
    \label{delta_rho_phi}
\end{equation}
The constant $\alpha$ is defined in (\ref{alphadef}) as
\begin{equation}
    \alpha \equiv \bra{\psi} \frac{\delta \rho_\phi}{\lambda^2} \ket{\psi}.
    \label{alpha'}
\end{equation}
Expanding out the trace in (\ref{delta_rho_phi}) gives (schematically)
\begin{equation}
    K(t,t') \equiv \Tr_A[[V(t), [V(t'), \rho_0]]] = gg' \left( \phi \phi'\rho_\psi R  - \phi \rho_\psi \phi' S - \phi' \rho_\psi \phi R+ \rho_\psi \phi' \phi S \right),
\end{equation}
where $R=\Tr_A(\rho_A X X')$ and $S = \Tr_A(\rho_A X' X)$. Then, since $\rho_\psi = \ket{\psi} \bra{\psi}$,
\begin{equation}
    \bra{\psi} K(t,t') \ket{\psi} = gg'\left[ R \bra{\psi} \phi \phi' \ket{\psi} - S \expval{\phi} \expval{\phi'}  - R \expval{\phi} \expval{\phi'} + S \bra{\psi} \phi' \phi \ket{\psi}\right].
\end{equation}
Since $\expval{\phi} = \expval{\phi'} = 0$, we obtain
\begin{equation}
    \alpha = - \int dt dt' \theta(t-t') g(t) g(t') [R W_\psi(t-t') + S W_\psi(t'-t)].
    \label{alpha}
\end{equation}
Note that $W_\psi(t'-t) = W_\psi(t-t')^*$ and $S=R^*$, so the second term in (\ref{alpha}) is the complex conjugate of the first, and one can write (\ref{alpha}) as
\begin{equation}
    \alpha = - 2 \int dt dt' \theta(t-t') g(t) g(t') \Re{R W_\psi(t-t')} \leq 0.
    \label{alphafinal}
\end{equation}

\section{The mutual information}
From the previous sections we know that the detector entropy change is
\begin{equation}
    \Delta S_A = \beta \langle Q \rangle + O(\lambda^4),
\end{equation}
while the field entropy change is
\begin{equation}
    \Delta S_\phi = \lambda^2 \alpha (\log \lambda^2 - 1) - \lambda^2 \sum_{i \geq 1} \mu_i \log \mu_i + O(\lambda^4).
\end{equation}
But from (\ref{entropy_conserved}) we know that the entropy of the full state is conserved. This can only be if
\begin{equation}
    S_{\text{tot}} = S_\phi + S_A - I_{\phi,A},
\end{equation}
where $I_{\phi,A}$ is the mutual information between the field and the detector. Initially, we have a product state $\rho_0$, which has zero mutual information, therefore
\begin{equation}
    I_{\phi,A} = \Delta S_\phi + \Delta S_A.
\end{equation}
In other words, the entropy of the subsystems increases by exactly the amount of correlations generated between them.

What does this mean? For a closed quantum system with unitary evolution, the von Neumann entropy is strictly constant. So if we insist that the entropy of the universe is the fine-grained quantity $S_{\text{tot}}(\rho)$ then the second law is trivial: the total fine-grained entropy does not change. 

The nontrivial second law is not about $S_{\text{tot}}$. It is about coarse-grained entropies, where we ignore some information. For example, if one looks at $S_\phi + S_A$ and ignores the mutual information term $I_{\phi,A}$, then the second law can be formulated as
\begin{equation}
    \Delta (S_\phi + S_A) = I_{\phi,A} \geq 0.
    \label{I>0}
\end{equation}
The apparent increase of entropy of the system together with the environment is the price of throwing away the correlation information between them. To leading order in $\lambda$, we can write (\ref{I>0}) as
\begin{equation}
    \alpha \lambda^2 \log \lambda^2 + \beta \langle Q \rangle \geq 0  , 
    \label{inequality}
\end{equation}
where $\langle Q \rangle$ is given by (\ref{Qangle}), and $\alpha$ is given by (\ref{alphafinal}).

Now, let's try to evaluate the two terms in the equation above explicitly. The Heisenberg equations corresponding to (\ref{H_A}) are $\dot{X}_A=P_A$ and $\dot{P}_A=-\Omega^2X_A$. Hence
\begin{equation}
    X_A(t) = X_A(0) \cos(\Omega t) + \frac{P_A(0)}{\Omega} \sin(\Omega t).
    \label{X_A(t)}
\end{equation}
Using $[X_A(0), P_A(0)]=i$, we find
\begin{equation}
    [X_A(t), X_A(t')] = -\frac{i}{\Omega} \sin[\Omega(t-t')].
\end{equation}
By the definition (\ref{chi})
\begin{equation}
    \chi_A(t-t') = i \theta(t-t') \Tr_A\left[ \rho_A^{(\beta)}[X_A(t), X_A(t')]\right] = \theta(t-t') \frac{\sin[\Omega(t-t')]}{\Omega}.
    \label{susceptibility}
\end{equation}
Then
\begin{equation}
    \dot{\chi}_A(t-t') = \theta(t-t') \cos[\Omega(t-t')].
\end{equation}
Plugging $\chi_A$ and its derivative into (\ref{Qangle}) we obtain
\begin{equation}
    \langle Q \rangle
    = \lambda^2 \int dt dt'\,
    g(t)\,g(t')\,\theta(t-t')\,\cos[\Omega(t-t')]\,
    W_\psi(t-t')\,.
    \label{Q-chi-cos}
\end{equation}
Now decompose $W_\psi(t-t')$ in its symmetric and antisymmetric parts: 
\begin{equation}
    W_\psi(t-t') = W_S(t-t') + iW_A(t-t'),
    \label{decomposition}
\end{equation}
with 
\begin{equation}
    W_S(t-t') = \frac{1}{2} \expval{\{\phi(t),\phi(t')\}}_\psi = \frac{1}{2} \expval{\phi(t) \phi(t') + \phi(t') \phi(t)}_\psi,
    \label{W_S}
\end{equation}
and 
\begin{equation}
    W_A(t-t') = \frac{1}{2i}  \expval{[\phi(t),\phi(t')]}_\psi = \frac{1}{2i} \expval{\phi(t) \phi(t') - \phi(t') \phi(t)}_\psi.
    \label{W_A}
\end{equation}
Note that because the cosine is an even function of $t-t'$, and $W_A$ is odd, the imaginary part drops out in the double integral, therefore we can as well write (\ref{Q-chi-cos}) as
\begin{equation}
    \langle Q \rangle
    = \lambda^2 \int dt dt' \theta(t-t')\,
    g(t)\,g(t')\,\cos[\Omega(t-t')]\,
    W_S(t-t')\,.
    \label{Qdec}
\end{equation}

In order to get a similar expression for (\ref{alphafinal}), we need to calculate $\Re{RW}$. For a harmonic oscillator at temperature $1/\beta$, the thermal averages are (established in appendix \ref{averages})
\begin{equation}
    \expval{X} = \expval{P} = 0, \quad \expval{X^2} = \frac{1}{2 \Omega}\coth{\frac{\beta \Omega}{2}}, \quad \expval{P^2} = \frac{\Omega}{2} \coth{\frac{\beta \Omega}{2}}, \quad \expval{XP} = - \expval{PX} = \frac{i}{2}.
    \label{expval}
\end{equation}
Given (\ref{X_A(t)}), we can write $X(t) X(t')$ schematically as 
\begin{equation}
    X^2 \cos(\Omega t) \cos(\Omega t') + \frac{XP}{\Omega} \cos(\Omega t) \sin(\Omega t') + \frac{PX}{\Omega} \sin(\Omega t) \cos(\Omega t') + \frac{P^2}{\Omega^2} \sin(\Omega t) \sin(\Omega t').
\end{equation}
Then, using (\ref{expval}),
\begin{equation}
    R = \expval{X(t)X(t')} = \frac{1}{2 \Omega} \coth{\frac{\beta \Omega}{2}} \cos{\Omega(t-t')} - \frac{i}{2\Omega} \sin{\Omega(t-t')}.
    \label{thermal2point}
\end{equation}
Note that (\ref{susceptibility}) and (\ref{thermal2point}) are related by the fluctuation-dissipation theorem, see appendix \ref{FDT}. In terms of the decomposition (\ref{decomposition}), $\Re{RW}$ is
\begin{equation}
    \Re{RW} = \frac{1}{2 \Omega} \coth{\frac{\beta \Omega}{2}} \cos{\Omega(t-t')} W_S(t-t') + \frac{1}{2\Omega} \sin{\Omega(t-t')} W_A(t-t').
\end{equation}
Therefore we can write (\ref{alphafinal}) as
\begin{equation}
    \alpha = - \frac{1}{\Omega} \coth{\frac{\beta \Omega}{2}} J_S - \frac{1}{\Omega} J_A,
\end{equation}
with
\begin{equation}
    \begin{aligned}
        J_S &= \int dt dt' \theta(t-t') g(t)g(t') \cos{\Omega(t-t')} W_S(t-t'), \\
        J_A &= \int dt dt' \theta(t-t') g(t)g(t') \sin{\Omega(t-t')} W_A(t-t').
        \label{integrals}
    \end{aligned}
\end{equation}
Similarly, (\ref{Qdec}) is
\begin{equation}
    \langle Q \rangle = \lambda^2 J_S.
\end{equation}
In terms of the integrals (\ref{integrals}), the inequality (\ref{inequality}) becomes
\begin{equation}
   J_S  \geq  \frac{\log \lambda^2}{\left( \Omega \beta -  \coth{\frac{\beta \Omega}{2}} \log \lambda^2 \right)} J_A.  
   \label{bound}
\end{equation}
Now recall that $J_A$ is defined in (\ref{integrals}) in terms of the field commutator (\ref{W_A}). Note also that for any free quantum field theory, $\expval{[\phi(x), \phi(y)]}_\psi = [\phi(x), \phi(y)] = i \Delta(x-y)$ as an operator identity, meaning that the commutator is a c-number, and proportional to the Pauli-Jordan function $\Delta(x-y)$. This means that $J_A$ is independent of the state $\psi$. All the state dependence of the two-point function $W_\psi$ lives in the symmetric part $W_S$, i.e. in $J_S$. Given that we assume $\lambda < 1$, $\log \lambda^2 < 0$, for every state $\psi$, the functional $J_S$ is bounded from below by a universal negative constant that depends only on the detector. 

Equivalently, since $\langle Q \rangle = \lambda^2 J_S$,
\begin{equation}
    \langle Q \rangle \geq \frac{\lambda^2 \log \lambda^2}{\left( \Omega \beta -  \coth{\frac{\beta \Omega}{2}} \log \lambda^2 \right)} J_A.
    \label{bound'}
\end{equation}
Physically, this means that the detector cannot cool below a certain threshold determined by its characteristic frequency, temperature, and the coupling with the field. The detector's heat is a non-local probe of the field's stress-energy tensor, therefore (\ref{bound'}) is also a bound on the field's energy density: the field cannot produce arbitrarily large negative energy, as measured by the detector.

\appendix
\section{Relation to the field's energy density}

We now explain how the detector heat $\langle Q\rangle$ is related to the smeared energy density $\mathcal{E}_g[\psi]$ defined in (\ref{Eg}). The key point is that both quantities are linear functionals of the same symmetric two-point function $W_S$, but with different kernels.

For a free scalar field, the renormalised energy density can be obtained by point splitting (\ref{T00}) on the worldline. Schematically,
\begin{equation}
    \langle T_{00}(t,\mathbf{0})\rangle_\psi
    = \frac{1}{2}\lim_{t'\to t}
    \Big[(\partial_t\partial_{t'} + \nabla\cdot\nabla')
     W_\psi(t,t') \Big]_{\mathbf{x}=0}.
\end{equation}
Along the inertial worldline $\mathbf{x}=\mathbf{0}$ the spatial derivatives can be traded for time derivatives, so that we may write
\begin{equation}
    \langle T_{00}(t,\mathbf{0})\rangle_\psi
    = \lim_{t'\to t}\,\partial_t \partial_{t'}W_\psi(t-t').
\end{equation}
Inserting this into (\ref{Eg}) gives
\begin{equation}
    \mathcal{E}_g[\psi] = \int dt g(t)^2 \lim_{t'\to t}\,\partial_t \partial_{t'}W_\psi(t-t'),
\end{equation}
or, equivalently,
\begin{equation}
    \mathcal{E}_g[\psi] = \int dt dt' \delta(t-t') g(t) g(t') \partial_t \partial_{t'}W_\psi(t-t').
\end{equation}
Now we integrate by parts to move the derivatives from $W_\psi$ to the smearing function $g(t)$. The boundary term vanishes because $g$ is smooth and compactly supported. Then, defining $F(t,t') \equiv \partial_t \partial_t' [g(t) g(t') \delta(t-t')]$,
\begin{equation}
    \mathcal{E}_g[\psi] = \int dt dt' F(t,t') W_\psi(t-t').
    \label{E_g}
\end{equation}
Since $F(t,t')$ is a symmetric function, only the (\ref{W_S}) piece of the decomposition (\ref{decomposition}) survives. Therefore, defining
\begin{equation}
    u = \frac{t+t'}{2}, \quad s = t - t',
\end{equation}
we can write (\ref{E_g}) as
\begin{equation}
    \mathcal{E}_g[\psi] = \int du ds F(u,s) W_S(s).
    \label{E_g(s)}
\end{equation}
If we additionally define the kernel $H(s)$ by integrating over the average time:
\begin{equation}
    H(s) \equiv \int du F(u,s),
\end{equation}
we can write (\ref{E_g(s)}) as
\begin{equation}
    \mathcal{E}_g[\psi] = \int ds H(s) W_S(s).
    \label{E_g_final}
\end{equation}
Starting from (\ref{Qdec}) and following the same steps that led to (\ref{E_g_final}) gives
\begin{equation}
    \langle Q \rangle = \lambda^2 \int ds G(s)\cos(\Omega s) W_S(s),
    \label{Q_final}
\end{equation}
with 
\begin{equation}
    G(s) \equiv \int du g(u+s/2) g(u-s/2).
\end{equation}
Both (\ref{E_g_final}) and (\ref{Q_final}) are linear functionals of the same symmetric two-point function $W_S(s)$, but with different kernels. Our entropy calculation yields a state independent lower bound on the detector functional $J_S$, hence on a particular average of $W_S$ defined by the kernel $G(s)\cos(\Omega s)$. Because $ \mathcal{E}_g[\psi]$ is another average of the same $W_S$ with a different kernel $H(s)$, this is naturally interpreted as a quantum inequality on a physically measurable, nonlocal probe of the field’s energy density: the field cannot produce arbitrarily large negative energy as seen by this thermometer.

\section{Thermal averages for the harmonic oscillator}
\label{averages}

In this appendix we derive the thermal expectation values (\ref{expval}).

The Hamiltonian is
\begin{equation}
    H = \frac{1}{2}(P^2 + \Omega^2 X^2).
\end{equation}
Introduce the standard ladder operators
\begin{equation}
    a = \sqrt{\frac{\Omega}{2}}\, X + \frac{i}{\sqrt{2\Omega}}\, P,
    \qquad
    a^\dagger = \sqrt{\frac{\Omega}{2}}\, X - \frac{i}{\sqrt{2\Omega}}\, P,
\end{equation}
which satisfy $[a,a^\dagger]=1$ and give the inverse relations
\begin{equation}
    X = \frac{1}{\sqrt{2\Omega}}(a + a^\dagger),
    \qquad
    P = - i \sqrt{\frac{\Omega}{2}}(a - a^\dagger).
\end{equation}
In terms of $a$ and $a^\dagger$, the Hamiltonian is
\begin{equation}
    H = \Omega\!\left(a^\dagger a + \frac{1}{2}\right).
\end{equation}
The thermal density matrix is diagonal in the number basis,
\begin{equation}
    \rho_\beta = \sum_{n=0}^{\infty} p_n \ket{n}\bra{n},
    \qquad
    p_n = (1 - e^{-\beta\Omega}) e^{-n\beta\Omega}.
\end{equation}
Consequently,
\begin{equation}
    \expval{a} = \expval{a^\dagger} = 0,
    \qquad
    \expval{a^\dagger a} = \bar n = \frac{1}{e^{\beta\Omega} - 1},
    \qquad
    \expval{a a^\dagger} = \bar n + 1,
    \qquad
    \expval{a^2} = \expval{a^{\dagger 2}} = 0.
\end{equation}
Using $X = (a + a^\dagger)/\sqrt{2\Omega}$, we obtain
\begin{equation}
    \expval{X} = \frac{1}{\sqrt{2\Omega}}(\expval{a} + \expval{a^\dagger}) = 0,
\end{equation}
and similarly $\expval{P}=0$. For the quadratic expectation values,
\begin{equation}
    X^2 = \frac{1}{2\Omega}(a^2 + a^{\dagger 2} + a a^\dagger + a^\dagger a),
\end{equation}
and therefore
\begin{equation}
    \expval{X^2}
    = \frac{1}{2\Omega}\big(\bar n + (\bar n + 1)\big)
    = \frac{1}{2\Omega}(2\bar n + 1)
    = \frac{1}{2\Omega}\coth\!\left(\frac{\beta\Omega}{2}\right),
\end{equation}
where we used
\begin{equation}
    2\bar n + 1 = \frac{e^{\beta\Omega}+1}{e^{\beta\Omega}-1}
    = \coth\!\left(\frac{\beta\Omega}{2}\right).
\end{equation}
A similar calculation for $P$ yields
\begin{equation}
    \expval{P^2}
    = \frac{\Omega}{2}(2\bar n + 1)
    = \frac{\Omega}{2}\coth\!\left(\frac{\beta\Omega}{2}\right).
\end{equation}
Finally, for the mixed correlator $XP$ we write
\begin{equation}
    XP + PX = -\, i (a^2 - a^{\dagger 2}),
\end{equation}
so its thermal expectation value vanishes because $\expval{a^2}=\expval{a^{\dagger 2}}=0$. Using also the canonical commutator $XP - PX = i$, we obtain
\begin{equation}
    \expval{XP} = \frac{i}{2},
    \qquad
    \expval{PX} = -\,\frac{i}{2}.
\end{equation}

\section{The fluctuation-dissipation theorem}
\label{FDT}
The detector susceptibility (\ref{susceptibility}) is
\begin{equation}
    \chi_A(t-t') = \theta(t-t') \frac{\sin[\Omega(t-t')]}{\Omega}.
    \label{chi_appendix}
\end{equation}
The thermal two-point function is given by (\ref{thermal2point}), so the symmetrized correlator is
\begin{equation}
    S_X(t-t') \equiv \frac{1}{2} \expval{\{X(t), X(t')\}} = \Re{R(t,t')} = \frac{1}{2 \Omega} \coth{\frac{\beta \Omega}{2}} \cos{\Omega(t-t')}.
    \label{S_X}
\end{equation}
The Fourier transform of (\ref{chi_appendix}) is
\begin{equation}
    \chi_A(\omega) = \frac{1}{\Omega} \int_0^\infty dt e^{i\omega t} \sin \Omega t = \frac{1}{2i\Omega} \int_0^\infty dt \left( e^{i(\omega+\Omega)t} - e^{i(\omega-\Omega)t}\right).
\end{equation}
The standard result
\begin{equation}
    \int_0^\infty dt e^{iat} = \pi \delta(a) + i \mathcal{P}\frac{1}{a},
\end{equation}
where $\mathcal{P}$ denotes the Cauchy principal value, gives
\begin{equation}
    \chi_A(\omega) = \frac{1}{2i\Omega} \left[ \pi(\delta(\omega+\Omega) - \delta(\omega-\Omega)) + i \left( \mathcal{P} \frac{1}{\omega+\Omega} - \mathcal{P} \frac{1}{\omega-\Omega}\right)\right].
\end{equation}
So
\begin{equation}
    \Im{\chi_A(\omega)} = \frac{\pi}{2\Omega} (\delta(\omega-\Omega) - \delta(\omega+\Omega)).
\end{equation}
The Fourier transform of (\ref{S_X}) is
\begin{equation}
    S_X(\omega) = \frac{1}{2 \Omega} \coth{\frac{\beta \Omega}{2}} \int_{-\infty}^{\infty} dt e^{i\omega t} \cos \Omega t = \frac{\pi}{2 \Omega} \coth{\frac{\beta \Omega}{2}}\left( \delta(\omega+\Omega) + \delta(\omega-\Omega) \right).
\end{equation}
Now look at the combination 
\begin{equation}
    \coth{\frac{\beta \omega}{2}} \Im{\chi_A(\omega)} = \coth{\frac{\beta \omega}{2}} \, \frac{\pi}{2\Omega} (\delta(\omega-\Omega) - \delta(\omega+\Omega)).
\end{equation}
Evaluating the distribution at the poles, and using the identity $\coth(-x) = - \coth(x)$ gives
\begin{equation}
    \coth{\frac{\beta \omega}{2}} \Im{\chi_A(\omega)} = \frac{\pi}{2\Omega}\coth{\frac{\beta \Omega}{2}}(\delta(\omega-\Omega) + \delta(\omega+\Omega)),
\end{equation}
namely
\begin{equation}
    S_X(\omega) = \coth{\frac{\beta \omega}{2}} \Im{\chi_A(\omega)}.
\end{equation}
The above relation is the fluctuation-dissipation theorem.

\section*{Acknowledgement}

I would like to thank Tony Rothman for posing the question that inspired this work.

\section*{Declarations}

\subsection*{Funding}  
No funding was received for this study.

\subsection*{Competing interests}  
The author declares no conflicts of interest.

\subsection*{Authors’ contributions}  
Single-author paper.

\subsection*{Code and data availability}
No new data or code were generated or used in this study.

\end{document}